# Mobile Application Threats and Security

Mark Miller, Shamimara Lasker, Michael Brannon
Information Technology
Georgia Southern University
Statesboro, GA USA

Dr. Tim Mirzoev
Information Technology
Georgia Southern University
Statesboro, GA USA

Abstract— the movement to mobile computing solutions provides flexibility to different users whether it is a business user, a student, or even providing entertainment to children and adults of all ages. Due to these emerging technologies mobile users are unable to safeguard private information in a very effective way and cybercrimes are increasing day by day. This manuscript will focus on security vulnerabilities in the mobile computing industry, especially focusing on tablets and smart phones. This study will dive into current security threats for the Android & Apple iOS market, exposing security risks and threats that the novice or average user may not be aware of. The purpose of this study is to analyze current security risks and threats, and provide solutions that may be deployed to protect against such threats.

Keywords- mobile security; mobile threats; Android; Apple iOS; malicious applications

## I. INTRODUCTION

Mobile computing is gaining vast popularity with much of the general public [12]. People are using mobile devices to access and store important data on a regular basis. Users record personal information such as names, address, phone number, and contacts [22]. In many mobile devices financial information is saved through applications or stored in browsing histories. Critical passwords and usernames are readily available to most users through mobile devices. Frequently, seemingly legitimate software applications, or apps, are malicious; it is not complicated to develop apps for some of the most popular mobile operating systems [12]. With information, mobile platform users are open to identity theft and account security risks. Mobile devices have become a big target for cyber criminals [6]. Companies that allow personal devices on corporate networks, 93 percent reported that when employees use smartphones, tablets, or other devices to work with the companies' information, customer information is at risk [6]. Application developers sacrifice security to save on time and effort, bringing products to market quicker by concentrating on the application's content [7]. The number of new vulnerabilities in mobile operating systems jumped 42 percent between 2009 and 2010 [12].

Mobile devices are now a pervasive part of everyday life in the United States. On a daily basis, millions of Americans use these devices to communicate and share information. Although mobile devices are quickly evolving due to technological breakthroughs, they are still highly vulnerable to a number of security attacks. According to the U.S. Government Accountability Office, security threats aimed at mobile devices have risen significantly [13]. The study conducted by the GAO in 2012 has found that the number of malicious software targeted at mobile devices increased from 14,000 to 40,000 in less than a year [13]. This issue is largely due to common vulnerabilities found in mobile applications.

One vulnerability found in mobile devices is the lack of security software [5]. Most devices do not use any form of security nor it comes preinstalled when purchased [5]. As a result, there is no protection against malicious attacks and spyware applications. There is also no way for users to detect these threats [5]. A common way for this is through harmful exploits that attack mobile devices through downloadable applications [5]. There are over 1 million applications available for download in both the iOS [2] and Android market [3]. Consumers can unknowingly download malware because it can effectively be disguised as a number of different applications such as a game, utility, security patch, and more [5]. Without any sort of security to detect a malicious application from one that is secure, there is higher risk of a security breach. It is because of these security factors; cyber criminals can easily access and steal user information without the user being aware.

Oversight by an application distributor can go a long way to ensure user security in mobile application development [21]. Penetration and other test methods are encouraged to identify security risk [25]. It is crucial for quality developers to have programming practices to be put in place to mitigate future security threats [21]. CompTIA realized the need for an industry wide certification focused on security in mobile application development [21]. The CompTIA Mobile App Secruity+ offers developers a certification in security awareness in mobile application programming [21]; however this does not address the immediate security needs for consumers.

Physical and software based security measures for mobile devices are some ways to mitigate security risks that are inherent in mobile computing [12]. Mobile device users have no excuse for having at risk devices. Many of the well-known security software developers now have mobile security apps available also [1]. Symantec survey of 12,704 respondents in 24 nations found that only 16% installed the most up-to-date security on their devices [1].

While mobile applications have enabled users to do work in a more convenient and efficient manner versus using slower loading mobile websites [14], the industry is realizing that many mobile applications are rushed through development, sacrificing security along the way [7]. There has been a recent movement to better regulate mobile application development as it relates to security through certification [21], but this only helps to protect mobile device users moving forward. The fact, that many applications are already in consumer markets that contain security vulnerabilities [13]; consumers must know how to protect themselves against security threats. This protection can be provided through an array of well-known security software providers, providing protection against malicious attacks that occur due to security exploits in the poorly developed mobile applications [1]. The education of end-users provides an insight to the types of threats while using mobile devices and mobile apps, along with the importance of utilizing security software on mobile devices. This study provides security analytics related to iOS and Android mobile users.

II. SUPPORTING DOCUMENTATION

Smartphones and tablets have been in wide use since the late 2000s. The introduction of the iPhone and the HTC Dream in 2007 and 2008 respectively allowed early mobile users to have the functionality of a computer in a compact hand carried device [16]. The iOS and Android operating systems made mobile computing a reality for the general population. The popularity of mobile computing caused cyber criminals to investigate ways of exploiting the new computing platform for criminal gain. The first attempts at exploiting the mobile computing platform were the use of techniques that criminals leveraged to exploit traditional systems, such as desktop and laptops that run Windows XP or Windows 7 [23]. Application development on both iOS and Android mobile devices created a new avenue for cyber criminals to exploit to gain information and access to mobile operating systems. Applications for mobile devices have security vulnerabilities in place in the code that the program is built from [8]. Some cyber criminals will write programs that are meant to look and act as a harmless and useful application to encourage victims to download the program. After malicious programs are installed these programs then send the sensitive information to the cyber-criminal to be used for identity theft [4]. The newest cyber intrusion techniques to target mobile devices focus on monitoring aspects of the phone that many experts consider to be low security threats. These applications monitor memory allocation to figure out which sites the victims are visiting and gyroscopes and accelerometers to determine which characters are being used on a keyboard. The information can be used to gather sensitive information that the victim enters into the mobile device [24].

Many mobile users are unaware of the threats that mobile devices can pose to information security. Mobile device users need be educated about measures that are necessary to keep the devices and the information they contain secure [23]. Like traditional systems mobile device security is always being challenged by cyber criminals with new and evolved methods to get or modify the information held in the devices [24]. The Security risk is a hazard to personal information as well as corporate information. Companies that have poor mobile security practices are opening the business to possible theft and information loss through unsecure mobile device [4]. Many government agencies have come to recognize the threat to sensitive information. These Agencies have either banned mobile devices or have allowed only agency acquired and maintained mobile device to be used in the line of duty to protect the information from being compromised [24].

Installing and deploying a security application on mobile devices can provide various protections against ever increasing threats. Several security applications have been developed to provide protection for personal users as well as users of an enterprise environment, assisting administrators in corporate or government environments. In the current market, some solutions are developed as a multi-device solution, such as Symantec's Norton 360 that covers PCs, Macs and Android mobile devices [9]; while others focus specifically on mobile devices with iOS and/or Android mobile operating systems in mind [11].

Norton 360 offers protection for personal users and in respect to mobile devices; it is only available for Android OS [9], while Fixmo's SafeWatch and Mocana's Mobile App Protection (MAP) suite is available for both iOS and Android devices [10] [20]. Fixmo's SafeWatch is also geared more towards the personal user but is available for enterprise deployment [10], while Mocana's MAP solution is strictly geared as an enterprise solution, allowing enterprise administrators the ability to remotely administer iOS and Android devices, regardless if they are a 'bring your own device' (BYOD) or a company owned device [20]. Other noteworthy differences between these security applications is SafeWatch's ability to transfer content between iOS and Android devices [10], MAP's ability to protect jailbroken iOS devices through app wrapping [20], and Norton 360's ability to provide parental controls [9]. For convenience, a more complete summary of features is provided in Table 1.

There have been many attempts to find a solution or at least decrease the rising number of mobile security threats. However, these efforts need to be more effective as many of them are unsuccessful or need great improvement. Google's Bouncer, a security program introduced February of 2012, is an example. Internet security company RiskIQ conducted a research which found a near 400% increase in the number of malicious apps on Google Play from 2011 to 2013 [18]. In order to protect Android users from these threats, Bouncer was designed to quietly scan apps for suspicious behavior and possible threats during the uploading stage to the market. Google has reported that as a result, the Bouncer program was responsible for reducing 40% of the number of harmful apps found in the Google Play [19].

TABLE I. SECURITY APPLICATIONS

| Summary of Features | Norton 360 | SafeWatch | MAP |
|---|---|---|---|
| Available for iOS | | ✓ | ✓ |
| Available for Android | ✓ | ✓ | ✓ |
| Protection for Viruses | ✓ | ✓ | |
| Protection for Fraudulent Websites | ✓ | ✓ | |
| Prevents eaves dropping | ✓ | | |
| Protection for Spam text | ✓ | ✓ | |
| Prevention of cybercriminal control of device | ✓ | | |
| Tracks Stolen/Lost Devices | ✓ | | |
| Remote Backup | ✓ | ✓ | |
| Remotely Lock | ✓ | ✓ | |
| Remotely Erase | ✓ | ✓ | ✓ |
| Remotely Take Photos | ✓ | ✓ | |
| Blocking calls/text | ✓ | ✓ | |
| Parental Controls | ✓ | | |
| Enterprise Solution | | ✓ | ✓ |
| Transfer content between iOS & Android | | ✓ | |
| Single Master Password for digital wallet & password vault | | ✓ | ✓ |
| Enterprise deployment of Apps on BYOD | | | ✓ |
| App protection through App wrapping approach | | | ✓ |
| Expanded VPN Capabilities | | | ✓ |
| Jailbroken iOS protection | | | ✓ |
| Browser configuration via Enterprise Admin. | | | ✓ |

Although this is a considerable improvement in protecting Android users, Bouncer is not invincible against all security threats. Dr. Oberheide and Dr. Miller, two security researchers, have found that Bouncer can be fingerprinted [17]. Among other findings, they have also revealed that Bouncer only scans an application for 5 minutes and only apps that behave suspiciously during the scan will be caught. It is not difficult to assume that malicious software will take advantage of these flaws. For example, malware designers can program submitted apps to behave nonthreateningly during Bouncer's security scan. Once the application is successfully installed onto the user's mobile device, it can then start to run the malware program as intended.

Overall, Bouncer can indeed stop harmful programs from reaching Google Play, thus preventing users from facing security threats. The problem lies with the fact that this security check can be easily evaded. Even if Google reprogramed Bouncer so that it is more effective in detecting harmful software, today's malicious programs are constantly evolving and finding other ways to bypass security. Mobile security threats are still greatly on the rise. Therefore, it is essential that programmers design better and safer software that will protect users and the information stored on their mobile devices.

III. TEST CASE PROCEDURE

To demonstrate the need for security software/applications on mobile devices a test installation of a known intrusive application was conducted on two android based platforms with a factory reset done between each install to ensure a clean environment. The first device tested was a first generation Amazon Kindle Fire. The Kindle Fire was equipped with the Android 2.3.3 Gingerbread operating system. The second device tested was the HTC One V Smartphone. The HTC One is equipped with the Android 4.0.3 Ice Cream Sandwich Operating System. Neither device came with any security applications preloaded to protect the devices. The application used to infect the devices in both the unprotected and protected stages was call Hungry Dino. This application could be found on both the Google Play store and the Amazon App store. The first security software that is used in the secure phase of testing is Trend Micro Maximum Security Titanium. Trend micro products are common off-the-shelf computer and mobile security software. The second security software that was used in the secure phase of testing was the Dr. Web application. The Dr. Web application can be found on both the Google Play store and the Amazon App Store. Each device was set to a factory state between each test to ensure that there were no extenuating circumstances that would affect results of the testing.

TABLE II. TESTED HARDWARE

| Hardware | CPU | RAM | OS | Wi-Fi | Internal Storage |
|---|---|---|---|---|---|
| Amazon Kindle Fire | Dual Core TI, 1.0 GHz | 512 MB | Android 2.3.3 | 802.11 b/g/n | 8 GB |
| HTC One V | Single core, 1.0 GHz | 512 MB | Android 4.0.3 | 802.11 b/g/n | 4 GB |

TABLE III. SECURITY SOFTWARE

| Software | Anti-Virus | Anti-Spam | Anti-Theft | Security Scan | URL Filtering |
|---|---|---|---|---|---|
| Trend Micro Maximum Security | ✓ | ✓ | ✓ | ✓ | ✓ |
| Dr. Web Anti-Virus for Android | ✓ | ✓ | ✓ | ✓ | ✓ |

The research conducted was based on the use of two devices, first generation Kindle Fire 1st Generation tablet and an HTC One V smartphone. Through these devices, the Hungry Dino app will be installed and used to infect both devices while security software Dr. Web and Trend Micro Internet Security Titanium will serve as a solution against the virus. The employed method proved to be successful due to the effectiveness of both security software and its ability to identify and remove malicious software from the devices. Both security programs are easily available and have been proven be successful against malicious software. The Dr. Web application can be found in the Google Play store where it has been downloaded by over 35 million users and has an average rating of 4.6 out of 5 [15]. Trend Micro Internet Security Titanium can be bought and downloaded from several locations and websites. For better protection, this security software also offers Android users extra features such as locating a device if it is lost or stolen and backing up stored data [26]. Both of the devices tested in this research used Android Operating systems which made them compatible to the security programs. Once the devices became infected with a virus, Dr. Web and Trend Micro Internet Security Titanium were able to immediately block the harmful software, or these anti-virus programs were be able to scan the full system, detect the malicious software, provide quarantine for detected threats, and offer users better protection from other dangerous applications. Android users will find that through these mobile security programs, they are far more protected from all types of malware programs.

In order to analyze and develop an understanding of each test result, data from each experiment was collected by observing and documenting adverse effects that the malicious app had on each device without a security application installed. After the installation of each security application, screen notifications were observed from both devices to determine the security application's ability to identify the threat. Other results recorded were the security application's ability to remove the threat from the device, or whether it only identified the threat and required manual removal by the user. After collecting the results of these four experiments, analysis was done by compiling results in the form of a table, thus providing a clear understanding of how each security application dealt with the same infectious mobile application on each device.

## IV. TEST CASE RESULTS

The first test observed the document reader application on the Amazon Kindle Fire with the Wi-Fi connection activated in the factory reset state. In this test 20 pages were navigated through in a 15 second time period. The HTC One V browser was activated in the factory reset state. In this test 5 pages of the walmart.com website were navigated in a 15 second period. This set of testing allowed for a baseline of normal operation under factory settings for both devices. In the second round of testing the factory reset of both devices, The Hungry Dino application was downloaded and installed on both devices. The internet connection was activated on both devices and the function testing for each device was administered the same way as the previous round of testing. In the second test the Amazon Kindle Fire lagged significantly in the reader test. The 20 page navigation took 40 seconds to complete because the Hungry Dino Application was using the CPU, RAM, and Wi-Fi connection to gather device information and send it to an outside source. On the HTC One V walmart.com 5 page site navigations took 5 minute to complete because the Hungry Dino application was utilizing the cell data connection, RAM, and CPU to collect and send device information to an outside source. The third round of testing both devices were reset to factory setting and The Dr. Web security application was downloaded installed and updated on both devices. The Hungry Dino application was then downloaded and installed on both devices. The Dr. Web application then scanned the Hungry Dino application and recognized the malicious code in the application and activated a warning to remove the Hungry Dino application. The remove application button was selected and the Hungry Dino application was deleted from the devices. The Amazon Kindle Fire reader test was 17 seconds to navigate the 20 page for this round of testing. The HTC One V navigated the 5 walmart.com pages in 16 seconds for this round of testing. In the fourth round of testing both devices were reset to factory settings. The Hungry Dino application was then downloaded and installed on each device. Then Trend Micro Internet Security was applied to each device and updated. The reader test was conducted on the Kindle Fire with a time of 45 second to navigate the 20 pages recorded. The website navigation test on the HTC One showed a time of 4 minutes 20 seconds. After these tests the security scan in the Trend Micro software was run on each device. During the scan the security software identified and removed the Hungry Dino application. The reader test was performed on the Kindle Fire with a time of 15 seconds for the 20 pages. The web browsing test was performed on the HTC One with a time of 20 seconds for the 5 pages.

TABLE IV. BASELINE RESULTS VS. POST INFECTION RESULTS

| Hardware | Pages Navigated (n) | Baseline Elapsed Total Time (sec) (No Infection) | Post Infection Elapsed Total Time (sec) (Infected with Hungry Dino App.) |
|---|---|---|---|
| Amazon Kindle Fire | 20 | 15 | 40 |
| HTC One V | 5 | 15 | 300 |

The results of the testing showed just how easy it is to accidently download and install a malicious program that can go about its tasks with little to no knowledge of the mobile user without close monitoring of the lag the mobile user might not realize that the malicious application is working in the background. The installation of security applications on the mobile devices recognized the malicious applications either immediately after installation or after the first security scan. Unlike traditional desktop and notebook computers mobile devices rarely have built in security measures to protect the system leaving them vulnerable to malicious software. These test demonstrated the need for mobile security and the importance of mobile security education for all mobile users.

## V. CONCLUSION

The purpose of the study was to make the reader aware of the security vulnerabilities of mobile devices. The research has shown that mobile devices are not equipped with standard security software. Mobile users that fail to take proper precautions, such as installing security applications, can unknowingly install malicious programs on their mobile devices and be subjected to theft of information or device performance degradation. Without proper security software the malicious programs can easily go undetected. Quite often performance degradations are discounted as software glitches or Internet connectivity problems. The applications tested in this case identified the malicious programs during install or after the first security scan when properly updated.

This study is meant to show the vulnerabilities inherent in many mobile devices in order to educate users to real security threats that many mobile users either ignore or are unaware of. Educating the average mobile device user is the first step to preventing security problems in mobile devices. With this awareness mobile users can take precautions to ensure the safety of information that is held on mobile platforms, by installing security applications to detect threats from malicious applications.

The research shows that security applications need to be integrated into mobile systems during the development period. The mobile device industry standards for device security should be raised to ensure that mobile users are protected from cyber-criminals and the software they use to carry out their criminal activities. Installing a security scan application onto each mobile device would make mobile users aware of malicious applications acquired when downloading applications. Information security needs to be taken more seriously by mobile device users and mobile device manufacturers.